\documentclass[11pt]{article}
\usepackage{graphicx,caption2,subfigure}
\makeatletter
\newcommand{\singlespacing}{\let\CS=\@currsize\renewcommand{\baselinestretch}{1.0}\tiny\CS}
\newcommand{\doublespacing}{\let\CS=\@currsize\renewcommand{\baselinestretch}{1.5}\tiny\CS}

\begin{document}
\title {Understanding the New High Energy Data-sets Measured by BESS, CAPRICE and PAMELA on Antiproton Flux
and $\overline{P}/P$ Ratios.}
\author { Goutam
Sau$^1$\thanks{e-mail:sau$\_$goutam@yahoo.com},P.
Guptaroy$^2$\thanks{e-mail:gpradeepta@rediffmail.com},A.
Bhattacharya$^3$\thanks{e-mail:aparajita$\_$bh@yahoo.co.in}$\&$ S.
Bhattacharyya$^4$\thanks{e-mail: bsubrata@www.isical.ac.in
(Communicating Author).}\\
{\small $^1$ Beramara RamChandrapur High School,}\\
 {\small South 24-Pgs,743609(WB),India.}\\
 {\small $^2$ Department of Physics, Raghunathpur College,}\\
 {\small Raghunathpur-723133,Purulia, India.}\\
 {\small $^3$ Department of Physics,}\\
 {\small Jadavpur University, Kolkata- 700032, India.}\\
 {\small $^4$ Physics
and Applied Mathematics Unit(PAMU),}\\
 {\small Indian Statistical Institute, Kolkata - 700108, India.}}
\date{}
\maketitle
\bigskip
\begin{abstract}
Reliable data from the very recent high-precision measurements on
the antiproton fluxes and the antiproton-to-proton ratios by the
PAMELA Collaboration at relatively much higher energies are now
available. The results with regard to antiproton production
phenomena spring no special surprises; rather they are sharply in
contrast with and contradiction to the case of positron production
at the same energy-range. However, the totality of data on
antiproton production from the past experiments to the very recent
PAMELA outburst seem to be a very challenging exercise for
interpretation in terms of the secondary production mechanisms
alone, the galactic propagation model etc against the background of
the `dark-matter'-related controversy. In the present work we assume
the validity of the simple leaky box model, choose a simple particle
production model and attempt at providing a comprehensive
interpretation of the totality of data on both antiproton flux
measurements and the $\overline{P}/P$ ratio-values for the various
experiments ranging from BESS, CAPRICE to the latest PAMELA
experiment. With the assumption of no contribution from the exotic
sources to the antiproton production process, our model and the
method describe the totality of the measured data with a fair degree
of success.
 Cosmic rays (including sources, origin, acceleration, and interactions).
 Dark Matter (stellar, interstellar, galactic, and cosmological). 
\end{abstract}
\newpage
\doublespacing
\section{Introduction}
In recent times the cosmic ray(CR) antiproton and positron flux
measurements have assumed much importance, as they could signal,
according to a section of the astroparticle physicists, to the
indirect detection of dark matter. In fact, the tremendous surge of
interest on the very recent PAMELA experiments centres around this
expectation. It is a fact that the PAMELA experiment on antiproton
production\cite{Adriani1} is, so far, not that controversial as is
the case for the positron excess measurements by the same PAMELA
group\cite{Adriani2}. And in a previous publication we successfully
dealt with the positron-excess-problem with some non-standard points
of view\cite{Sau1}.
\par In the astroparticle sector the observed particles fall under
two categories : primaries and secondaries. The primaries are
considered to be those which are accelerated by astrophysical
objects of our galaxy and comprise of electrons, protons, other
light and heavier nuclei\cite{Delahaye1}. In the course of its
journey in the space these energetic primaries undergo the process
of spallation on the Interstellar Medium (ISM) mainly with the
components Hydrogen and Helium. Besides, they also suffer energy
losses by successive high energy interactions in the Earth's
atmosphere while passing through the earth. After their production
from the supernovae remnants, cosmic rays traverse and propagate in
the galactic turbulent magnetic field, experience some deflections
and finally enter the Earth. Both the spallation process and the
interactions of the primaries with the ISM give rise to the various
secondaries of which the antimatter particles, e.g., positrons,
antiprotons etc. constitute a considerable part. For antiproton
production spallation of secondaries too might play some role,
though for all practical purposes we will certainly neglect such
tertiary mode of production.
\par The concept of dark matter and its coupling to the Standard Model (SM)
sector allows some annihilation or decay chains which could be
additional sources of both matter and antimatter in equal measure.
But in the general background matter is much more abundant than
antimatter. So, it is quite understandable that the primary
component due to so-called Dark Matter has relatively better chances
to be detected\cite{Dar1} among antimatter cosmic rays. In a way,
this accounts for the increased importance of the antimatter studies
in the recent times.
\par In view of the importance of antiproton research in appreciating
the role of production and propagation of primary cosmic rays, we
have motivated ourselves here to find a compatibility of the various
data-sets by using the simple leaky box model (SLBM) for galactic
propagation and using a particular secondary antiproton production
model as described in some detail in the next Section. The
observations made and reported by various groups like,
BESS`95\cite{Orito1}, BESS`97\cite{Orito1}, BESS`99\cite{Asaoka1},
BESS`00\cite{Asaoka1}, BESS`02\cite{Haino1} etc.,
CAPRICE`98\cite{Boezio1} and PAMELA(2008)\cite{Adriani1} would here
be dealt with on the basis of the SLBM\cite{Cowsik3} alone which is
nowadays very much a text book matter, for which no further details
about it would be presented here.
\par In the present work we will assort the data on antiproton
flux measurements and the $\overline{P}/P$ ratios by BESS[1995 -
2002], CAPRICE[1998] and PAMELA groups and try to understand the
totality of data in a comprehensive manner with the clear emphasis
laid on the latest PAMELA results.

\section{Mechanism for Antiproton Production and the Expression for Invariant Cross Sections}
According to the secondary production model the low-$p_T$ (soft)
baryon-antibaryon secondaries are produced here through the decays
of (virtual) secondary pions of which proton-antiproton pairs
comprise nearly one third of the total. Bandyopadhyay and
Bhattacharyya\cite{Bandyopadhyay1} and Bandyopadhyay et
al\cite{Bandyopadhyay2} have worked out the details of the necessary
field-theoretic calculations based on Feynman diagrams and obtained
the following formulae for inclusive cross-sections at low-$p_T$
valid for moderately high to high energies and by the average
antiproton multiplicity
\begin{equation}
E\frac{d^3\sigma}{dp^3}|_{pp\longrightarrow\overline{p}X}\simeq1.87 \times exp[-7.38\frac{p{_T}^{2}+m^{2}_{\overline{p}}}{1-x}]exp[-5.08x]
\end{equation}
and
\begin{equation}
<n_{\overline{p}}>\simeq1.08 \times 10^{-2}S^{2/5}~~~~~~~~~~~~~~~~~~~~~~~~~~~~~~~ for \sqrt{S}\leq100GeV
\end{equation}
\begin{equation}
<n_{\overline{p}}>\simeq2 \times 10^{-2}S^{1/4}~~~~~~~~~~~~~~~~~~~~~~~~~~~~~~~ for \sqrt{S}>100GeV
\end{equation}
where $m_{\overline{p}}$ is the mass of the antiproton and
$n_{\overline{p}}$ is the measured antiproton multiplicity. With (2)
we get at $\sqrt{S}$=53GeV, $<n_{\overline{p}}>\simeq$0.2 for both
the formulae.
\par The points of emphasis about this model are : i) It gives dynamically
a unified picture of both low- and large-transverse-momentum
phenomena and admits of no compartmentalization between soft and
hard production of particles which is an artifact from the dictates
of the Standard Model in particle physics. The only difference
between them, according to this model, is in kinematics and in one
additional feature of constituent rearrangement at high $p_T$. ii)
It explains the $\ll universality \gg$ property of high-energy
lepton-hadron, hadron-hadron, hadron-nucleus collisions and $e^+e^-$
reactions in a nice dynamical and unambiguous way. iii) It
subscribes to the ideas of jettiness of particle production at high
energies in the form of two-sided $\ll sprays \gg$ of sequential
arrays of hadrons. iv) It explains the by-now established
leading-particle effect (LPE) in high- and very-high-energy
collisions in a very satisfactory manner. v) Save and except a
single parametrization, wherein there is a degree of uncertainty,
there is no hand-inserted parameter in the model. The model proposes
a power law multiplicity for high-energy particle production and
introduce Feynman scaling violation in an inbuilt manner even for
relatively low-transverse-momentum region. For exceedingly low-$p_T$
region a logarithmic nature of multiplicity might work and Feynman
scaling might be valid under some stringent conditions and some
strict restrictions. Naturally, this region is very limited and only
a case for exception. vi) Last but not the least, another potential
success of the model lies in its ability to explain the very slow
rise of $K/\pi$ ratio emphasized\cite{Chou1,Alner1} even very
recently.
\section{Estimation of Antiproton Flux and the $\overline{P}/P$ Ratios}
This Section is divided into the following few subsections and the
undernoted sub-captions.
\subsection{Primary Spectra, Secondary Antiproton Production Model and the Working Formulae}
In actual evaluation the model dependence comes into picture for
getting values of d${\overline{\sigma}}$/dE which is related with
the inclusive cross-section in the following way\cite{Tan1} :
\begin{equation}
\frac{d\overline{\sigma}}{dE}=\frac{\pi}{p_L}\int(E\frac{d^3\sigma}{dp^3})_{pp\rightarrow\overline{p}X} ~ dp_T^2
\end{equation}
Here we take p$_L\simeq$E as the transverse momenta of the produced
secondary antiprotons is assumed to be small.
\par Inserting expressions (1), (2) and (3), our model-derived
formula for inclusive cross-section valid at moderately high
energies in eqn.(4) and integrating over $p_T$ with normal
approximations we get
\begin{equation}
\frac{d\overline{\sigma}}{dE}|_{p\rightarrow\overline{p}}\simeq0.496~exp[-5.08x]
\end{equation}
where we have used the low-transverse-momentum upper limit up to
$p_T$= 1 GeV/c. It must also be recalled that $p_L\simeq$E.
Bhattacharyya and Pal\cite{Bhattacharyya1,Bhattacharyya2} have
worked out that the antiproton-to-proton ratio is to be given
finally by
\begin{equation}
f_{\overline{p}}(E)=\frac{J_{\overline{p}}(E)}{J_p(E)}=\frac{2K\lambda_e(E)}{m_p}\int^{X_s}_0E\frac{d\sigma_p}{dE}X^{\gamma-1}dx
\end{equation}
where $J_p$ and $J_{\overline{p}}$ are the differential fluxes of
the primary protons and the secondary antiprotons ($m^{-2} sr^{-2}
s^{-1} GeV^{-1}$) respectively. K is the correction coefficient
taking into consideration the composition of the primary cosmic rays
and the interstellar gas, $\lambda_e(E)$ is the average path length
of antiprotons against escape ($g cm^{-2}$ as the unit), $m_p$ is
the mass of the proton (g as the unit), $E_p$ is the total energy of
the primary proton, $E_s$ is the integral lower limit relevant to
the production threshold of antiprotons, $\gamma$ is the integral
energy spectrum exponential of the primary protons and is the sole
quantity taken from cosmic-ray information and it is used
$\gamma$=1.75, X = E/$E_p$ and $X_s$ = $E_s/E_p$ (we took $X_s
\simeq$ -($m_pc^2$/E)+[$(m_pc^2/E)^2$+1]$^{1/2}$).
\par Usually, $J_p$ is expressed as
\begin{equation}
J_p(E_p) = J_0 E_p^{-(\gamma+1)}
\end{equation}
Now using eqn.(5), eqn.(7) in eqn.(6) we get
\begin{equation}
\frac{f_{\overline{p}}(E)}{K\lambda_{e}(E)}=\frac{2}{m_p}\int_0^{X_s}0.496 ~ exp[-5.08x]x^{\gamma-1}dx
\end{equation}
Here, we have always used K = 1.26, $\lambda_e$ = 5 $g cm^{-2}$ and
$m_p \simeq$ 1GeV.

\subsection{The Effect of Annihilation Channels : A Damping Effect}
At the relatively lower energy sides of ultra high energy
interactions, both proton-antiprotons might enter into some
annihilation reactions which would suppress the production ratio of
$\overline{P}/P$ in the main. In order to accommodate this
probability we assume a damping correction term for the
$\overline{P}/P$-ratios to be parameterized by $\sim \phi
\exp[-\alpha E_{\overline{P}}^\epsilon]$ where $\phi$, $\alpha$ and
$\epsilon$ are the chosen parameters and $E_{\overline{P}}$ is the
measured antiproton energy. We assume here that only the
ratio-values of $\overline{P}/P$ would suffer this diminutive change
as the annihilations involve both the protons and antiprotons,
though the individualized production of both protons and antiprotons
would be considered to remain unaffected. The used values of $\phi$,
$\alpha$ and $\epsilon$ are 0.55, 0.7 and 0.5 respectively.

\subsection{Choice of the Working Primary Proton Spectra}
It is to be observed that the value of
$f_{\overline{P}}(E)/K\lambda_e(E)$ is sensitive to the value
$\gamma$ and the value of $\gamma$ could be different in different
regions of the primary proton energy. But we have left out this
issue for the present with the acceptance of a specific value of
$\gamma$ = 1.75 throughout this entire work. In order to proceed we
have found the relationship between the secondary antiproton energy
and the primary proton energy to follow the nature depicted by
figure 1(a).
\par In figure 1(b), the median energy of primary protons is shown
as a function of the secondary antiproton energy. It was found
earlier that for antiproton energies of 3-9 GeV, which were relevant
to the experimental work performed uptil then, the median energies
of primary protons were about 25-80 GeV. The present energy-region
is somewhat higher. But we take the cue from Tan and Ng\cite{Tan1}
and proceed in a similar manner to draw the figures shown in figure
1(b) and figure 1(c) as described in the figure-captions in some
detail. Very carefully, we have chosen a modestly accurate primary
proton spectrum. Modifying Bhadwar et al\cite{Badhwar1} , we use
here
\begin{equation}
J_P(E_P) = 2 \times 10^5 E_P^{-2.75}
\end{equation}
where $J_P(E_P)$ is in protons $m^{-2} sr^{-2} s^{-1} GeV^{-1}$.
\par Using this spectrum and our invariant cross section formulae, we have
calculated the $f_{\overline{P}}(E)/K\lambda_e(E)$ curve as given in
figure 1(c).

\section{Results}
The final results have here been actually worked out on the basis of
the following two deduced expressions :
\begin{equation}
f'_{\overline{P}}(E_{\overline{P}}) = f_{\overline{p}}(E) J_P(E_P)
\end{equation}
and
\begin{equation}
R_{\overline{p}}(E_{\overline{P}})=\frac{J_{\overline{p}}(E)}{J_p(E)} \times  \phi
\exp[-\alpha E_{\overline{P}}^\epsilon] = \frac{2K\lambda_e(E)}{m_p}\int^{X_s}_0E\frac{d\sigma_p}{dE}X^{\gamma-1}dx \times \phi
\exp[-\alpha E_{\overline{P}}^\epsilon]
\end{equation}
The graphical plots drawn in figure 2 and figure 3 with expression
(10) describe the nature of relatively low-energy antiproton-data
measured by BESS, CAPRICE on antiproton flux. And the plot of
model-based $\overline{P}/P$ ratio-values based on expression (11)
are displayed in figure 4 and figure 5 against the data-background.
The used values of the parameters are shown in the adjoining table
(Table 1).
\par After delivering the results by graphical plots and the
necessary table(s), we would like to make in the following section
some crucial observations about some very exciting and interesting
recent works\cite{Blasi1,Blasi2,Ahlers1,Mertsch1} which appear to be
somewhat concurrent with our views on some aspects of astroparticle
physics-divisions. The final chapter of discussions and conclusions
drawn from the present work just follows thereafter.
\section{Some Contemporary and Novel Studies on PAMELA-Results : The Specific Features and a Few Comments}
In the very recent past Blasi\cite{Blasi1} proposed an ``alternative
and even simpler astrophysical explanation"\cite{Blasi2} for the
anomalous positron excess observed and reported by PAMELA
group\cite{Adriani2}. Of late, thereafter the same postulates (plus
mechanism) of hadronic production of secondaries in aged supernova
remnants (SNRs) and acceleration therein have been successfully
applied for calculating the $\overline{P}/P$ flux-ratio which has
been generically predicted to be of flattened nature in the
intermediate energy range under study now, and is expected to rise
very weakly in the superhigh energy domain. In fact, the initial
experimental observations on this ratio in the TeV-region
corroborate this predicted trend of the data. But confirmation of
this behaviour is still awaited by both the theorists and
experimentalists\cite{http1}.
\par The different probable sources for production of antimatter
components, like positrons, antiprotons etc are : dark matter
annihilation/decay, secondary production and supernova
remnants\cite{Lineros1}. These three are mutually independent of
each other, apparently with no known and perceptible overlapping or
intersection. The SNR components are transient sources, in which
most of the matter-antimatter particles are accelerated and injected
into the interstellar space in a very short time compared to cosmic
ray propagation scales. The difference with the two other sources
mentioned above is that those are rather inhomogeneous in the nearby
region to the solar system. Normally this kind of source is taken as
a smooth distribution across the whole galaxy due to the diffusive
propagation of cosmic rays. However, when sources are too close - on
time and distance - the distribution pattern changes and the
approximation is no longer valid.
\par However, that the secondary to primary ratios should rise with
energy, if secondaries are accelerated in the some spatial region as
the primaries, have been noted quite some time ago in the context of
cosmic ray acceleration in the interstellar medium. This model is
conservative since it invokes only processes that are expected to
occur in candidate cosmic ray sources, in particular supernova
remnants. Our way to distinguish it from the other models is to
compute the expected antiproton-to-proton ratio, which is
experimentally observed so far to be consistent with the standard
background\cite{Ahlers1,Mertsch1}. This finding is, in fact, in
agreement with the model by Blasi\cite{Blasi1}, more particularly by
Blasi and Serpico\cite{Blasi2}. It is to be noted that Blasi's
acceleration mechanism applies to a certain stage of the (old) SNR
evolution with some other provisos and conditionalities related to
magnetic field damping, time-dependence or time-independence etc.
However, till date, the model is found to be consistent with the
measurements.
\par The commonness in the approaches by us, Blasi [or Blasi and
Serpico] and Ahlers et al lie in the facts that (i) all of these
models are not based on the assumed/hypothetical existence of the
dark matter or, for that matter, decay/annihilation of dark matter
etc.; and (ii) there is no concern or consideration about the
existence of the pulsars and the role of the emissions from the
pulsars. In our mechanism, we confined ourselves only to the
`secondary' proton-antiprotons. Blasi, Blasi and Serpico opted
singularly for the third source, the SNRs, which we did not, at all,
reckon with. So these two approaches might be viewed as parallel to
each other. In future, it is quite probable that a combination of
these $\underline{two}$ contributions would be an indispensable
necessity to have a perfectly valid explanation for the measured
data at ultrahigh energies. In order to confirm/discard any model
whatsoever under consideration here , one needs to have very
reliable high-statistics data simultaneously on the individual
fluxes of both the protons, antiprotons and the ratio-values of the
antiprotons-to-protons as well.
\section{Concluding Remarks}
The entire data-sets measured by three distinctly separate groups at
somewhat different energy-ranges of the proton primaries have here
been described in an integrated manner and in a modestly
satisfactory way. In our calculations we have not taken into
consideration any exotic source of antiprotons. So, no specific
support to the concept of `dark matter' could be rendered by our
calculations and method; rather in a tacit manner the idea is
virtually disfavored. This is in contrast with the contentions of
Buchm$\ddot{u}$ller et al\cite{Buchmuller1}. Even the probable
emissions from the pulsars find no place in our approach. With
regard to all these exoticities, we obviously share the ideas of
both Blasi\cite{Blasi1} and Blasi and Serpico\cite{Blasi2}.
\par We have based our
calculations here on the Simple Leaky Box Model (SLBM). Even our
previous work on positron results\cite{Sau1} by PAMELA
group\cite{Adriani2} was also done with the same propagation model,
viz, SLBM. So the emphasis laid on the Nested Leaky Box Model (NLBM)
by Cowsik and Burch\cite{Cowsik1,Cowsik2} on understanding the
PAMELA-data related to the detection of positron excess is not
substantiated by our works. Besides, the fair agreement between
calculations and measurements provides a support to the secondary
antiproton production model and the choice of the nature of primary
proton spectrum put into use here. But, in our approach there are a
few chosen parameters which cannot be clearly accounted for right
now from the physical considerations. This definitely constitutes
some strong limitations to our claims on the successes in
interpreting the data-trends and/or data-character(s).
\begin{table}
\begin{center}
\begin{small}
\caption{Chosen values of the parameters}
\begin{tabular}{|c|c|c|c|c|c|c|}\hline
$\lambda_e(g cm^{-2})$ & $m_p(GeV)$ & $\gamma$ & $K$ & $\phi$ & $\alpha$ & $\epsilon$\\
\hline
$1.26$&$\simeq1.00$& $1.75$ & $5.00$&$0.55$&$0.70$ & $0.50$\\
\hline
\end{tabular}
\end{small}
\end{center}
\end{table}
\par \textbf{Acknowledgement}
\par The authors express their thankful gratitude to the honourable
referee for making an insightful comment and a constructive
suggestion for improvement on an earlier draft of the manuscript.

\newpage
\singlespacing

\newpage

\begin{figure}
\subfigure[]{
\begin{minipage}{.5\textwidth}
\centering
 \includegraphics[width=3.0in]{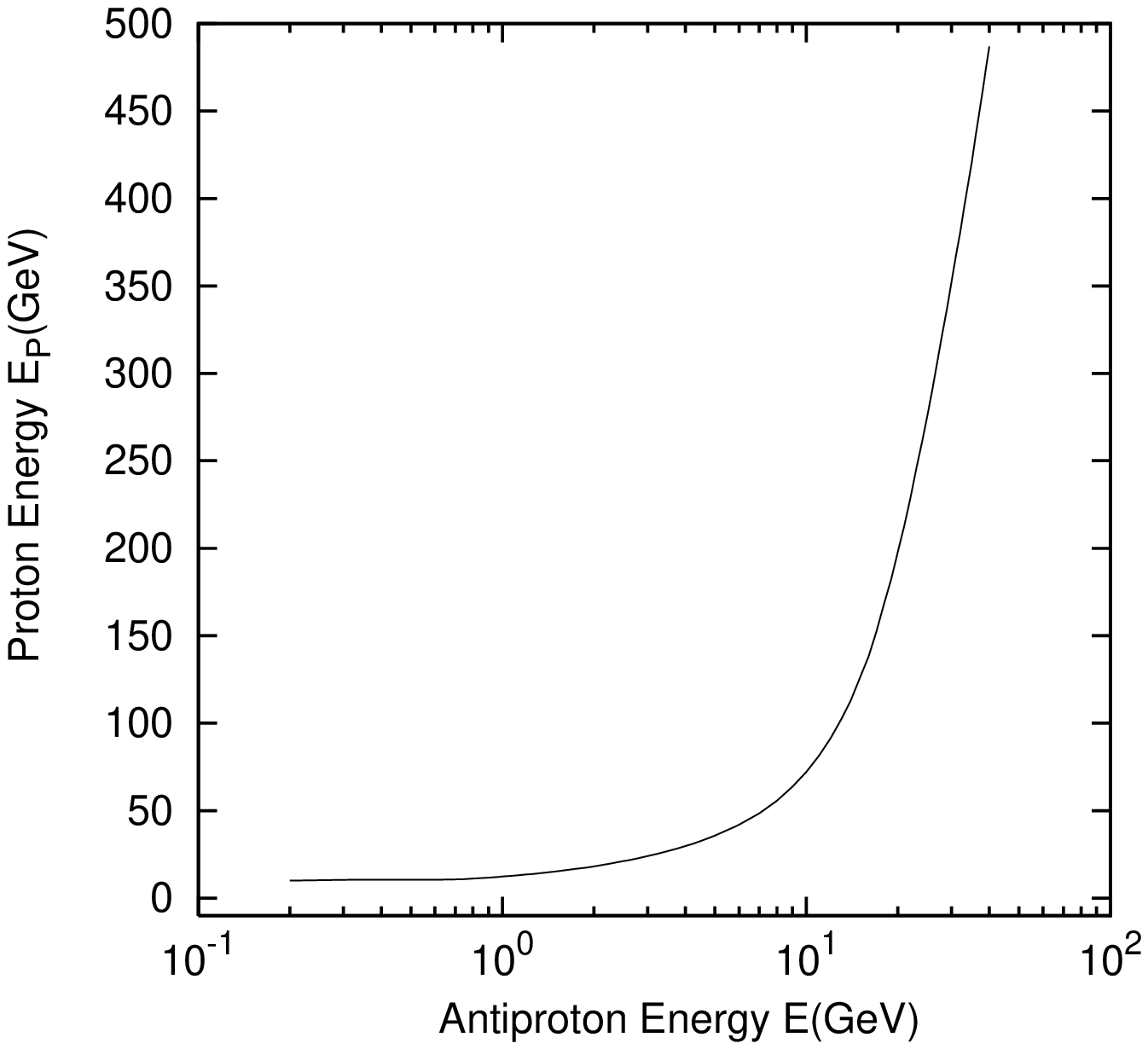}
\end{minipage}}%
\subfigure[]{
\begin{minipage}{0.5\textwidth}
  \centering
\includegraphics[width=3.0in]{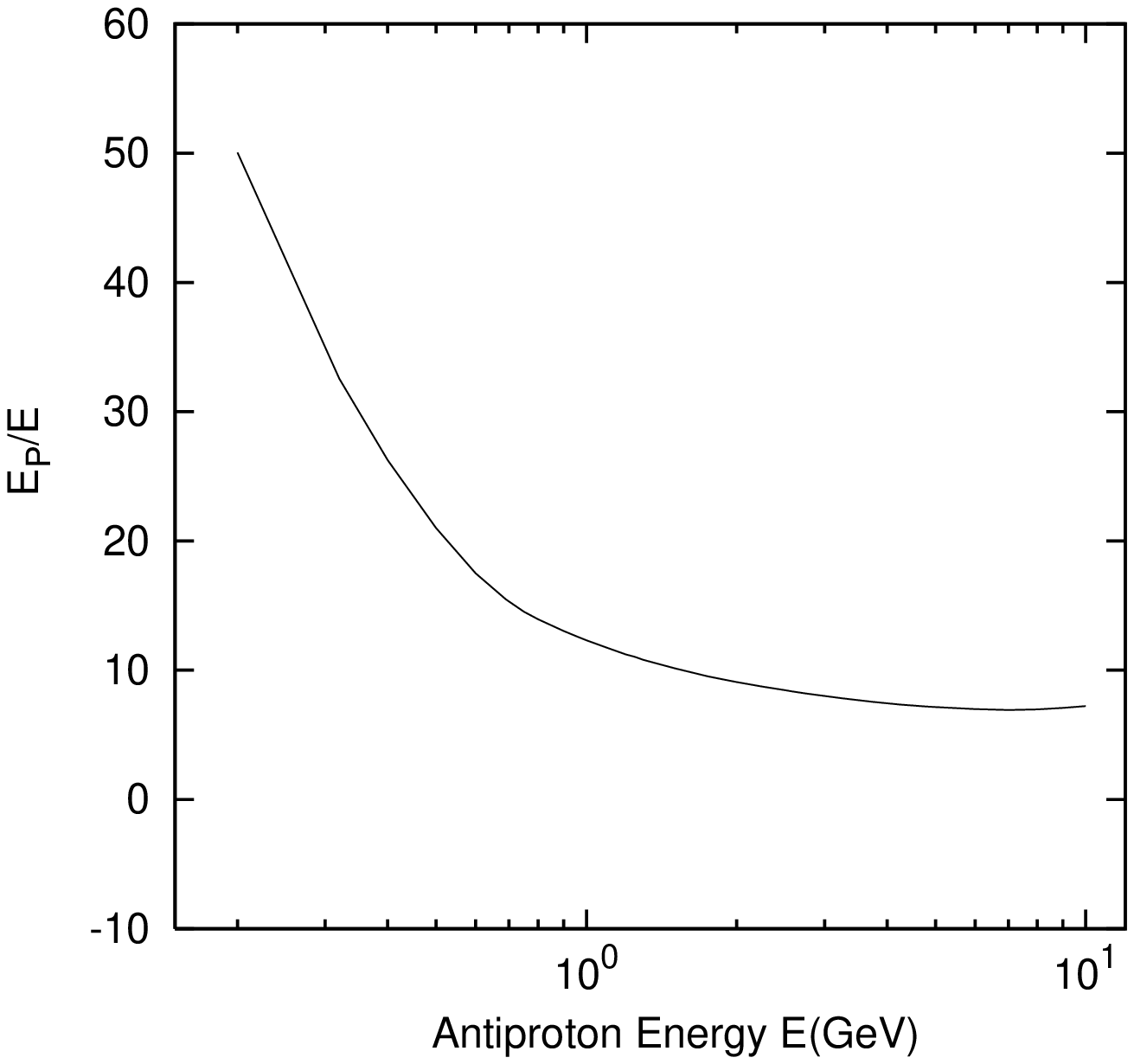}
\end{minipage}}%
\vspace{.01in} \subfigure[]{
\begin{minipage}{1\textwidth}
\centering
 \includegraphics[width=3.0in]{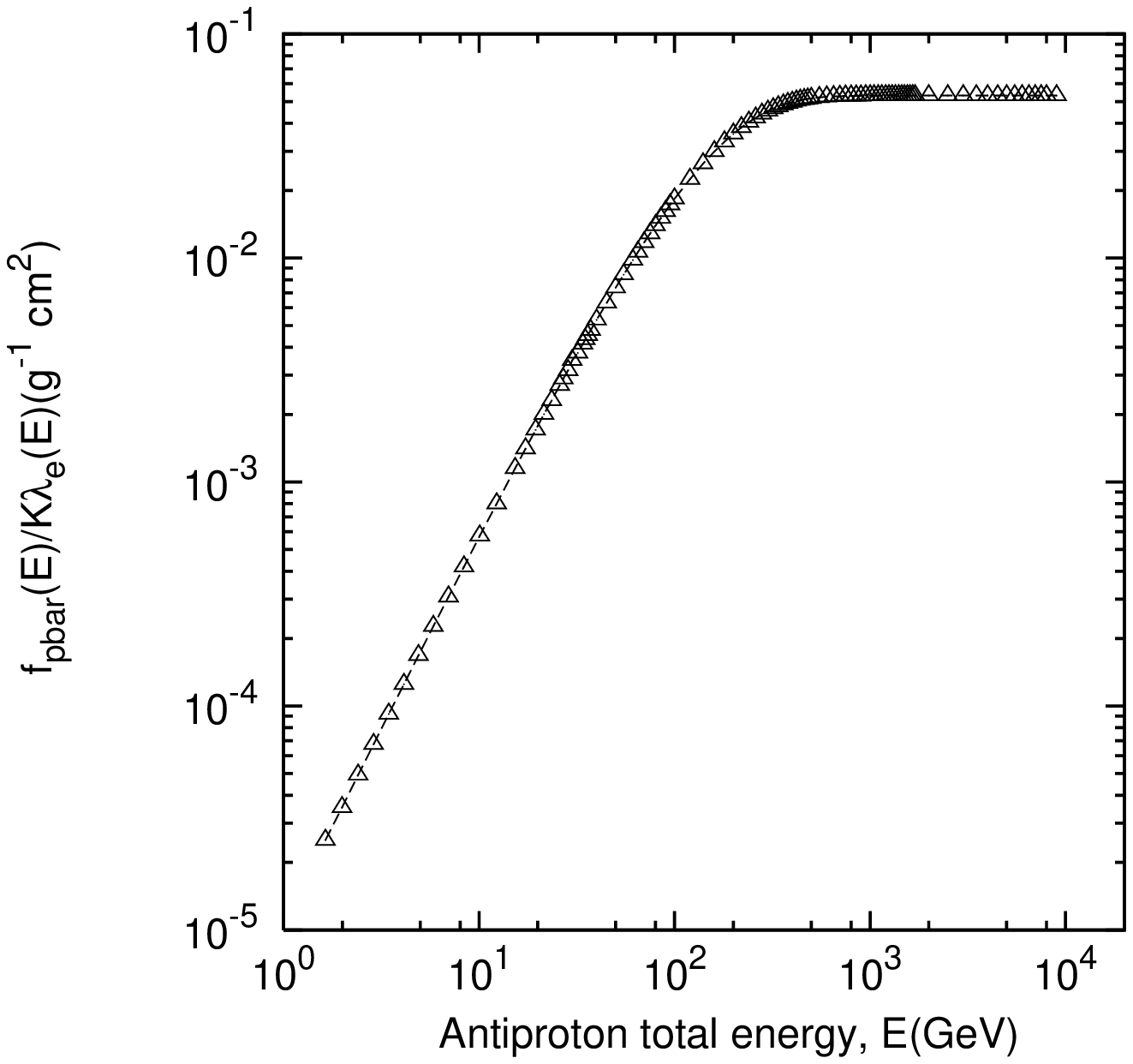}
\end{minipage}}%
\caption{(a) Plot of the relationship between the variation of antiproton energy (E) and primary proton energy ($E_P$),
 (b) Relation between primary proton energy ($E_P$) and the total antiproton energies (E); the plot has been done with
 $E_P/E$ as Y-axis and E-valus as X-axis, (c) The calculated antiproton to proton flux ratio (in terms of $K/\lambda_e(E)$)
against the antiproton energy E.}
\end{figure}

\begin{figure}
\subfigure[]{
\begin{minipage}{.5\textwidth}
\centering
\includegraphics[width=3.0in]{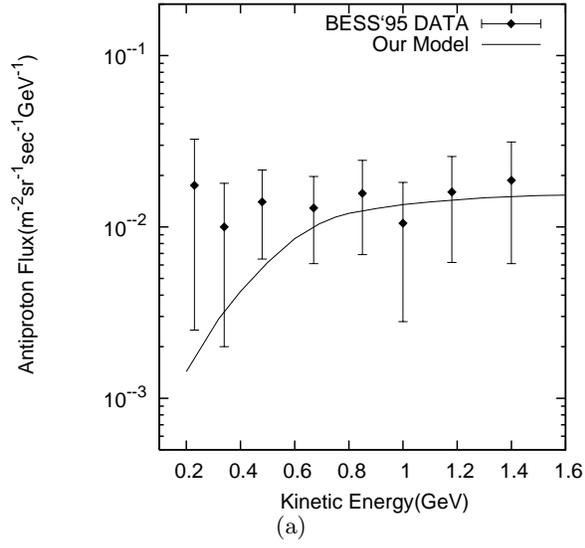}
\setcaptionwidth{2.5in}
\end{minipage}}%
\subfigure[]{
\begin{minipage}{0.5\textwidth}
\centering
 \includegraphics[width=3.0in]{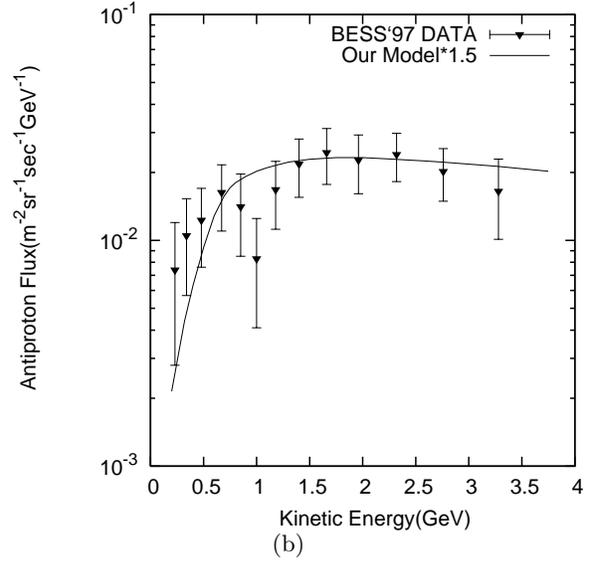}
 \end{minipage}}%
\vspace{0.01in} \subfigure[]{
\begin{minipage}{0.5\textwidth}
\centering
\includegraphics[width=3.0in]{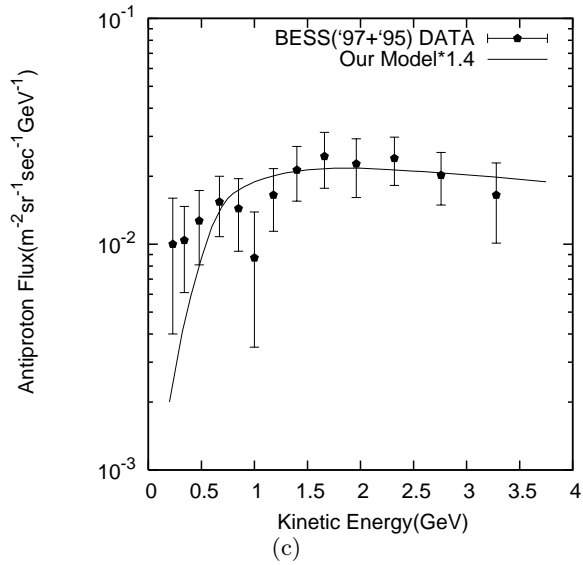}
\end{minipage}}%
\subfigure[]{
\begin{minipage}{.5\textwidth}
\centering
 \includegraphics[width=3.0in]{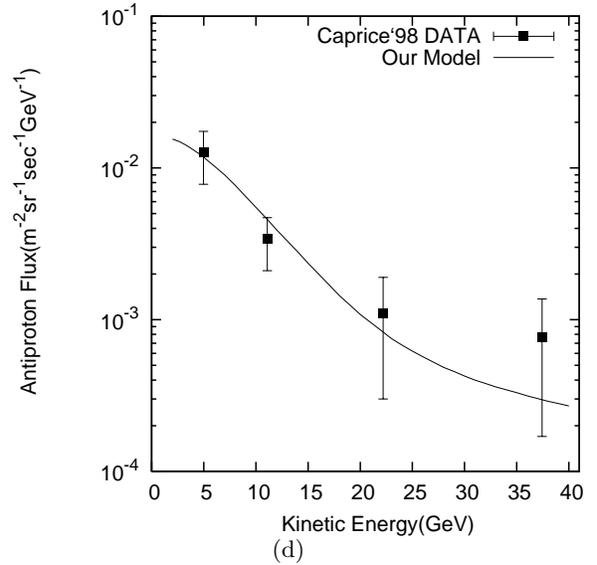}
 \end{minipage}}%
\caption{Antiproton fluxes at the top of the atmosphere
as measured by BESS`95[2(a)], BESS`97[2(b)], BESS(`97+`95)[2(c)] and Caprice`98[2(d)].
The solid lines represent the calculations based on our theoretical model (10).
 The experimental data are collected from Ref.\cite{Orito1,Boezio1}}
\end{figure}

\begin{figure}
\subfigure[]{
\begin{minipage}{.5\textwidth}
\centering
\includegraphics[width=3.0in]{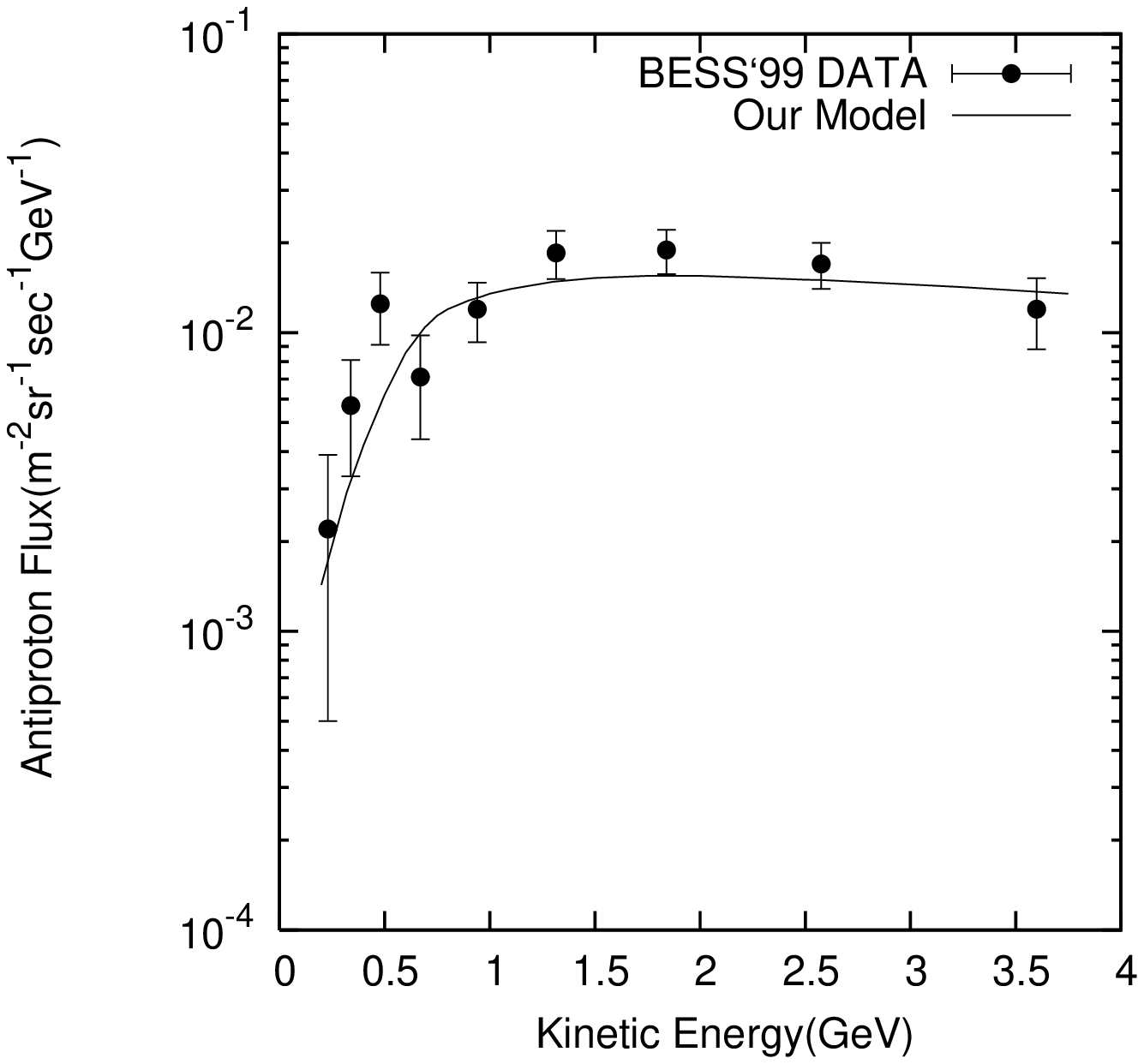}
\setcaptionwidth{2.5in}
\end{minipage}}%
\subfigure[]{
\begin{minipage}{0.5\textwidth}
\centering
 \includegraphics[width=3.0in]{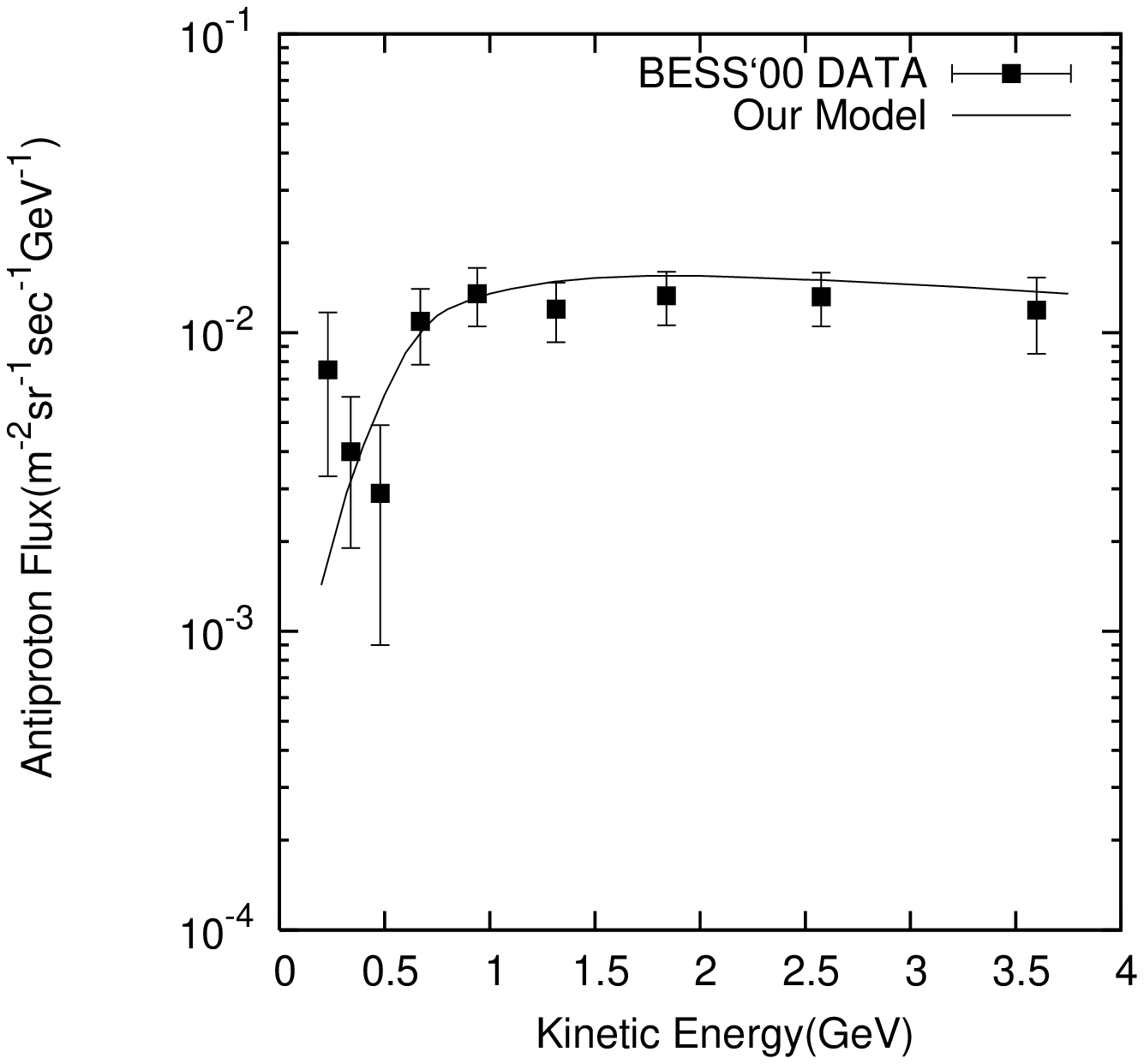}
 \end{minipage}}%
\vspace{0.01in} \subfigure[]{
\begin{minipage}{0.5\textwidth}
\centering
\includegraphics[width=3.0in]{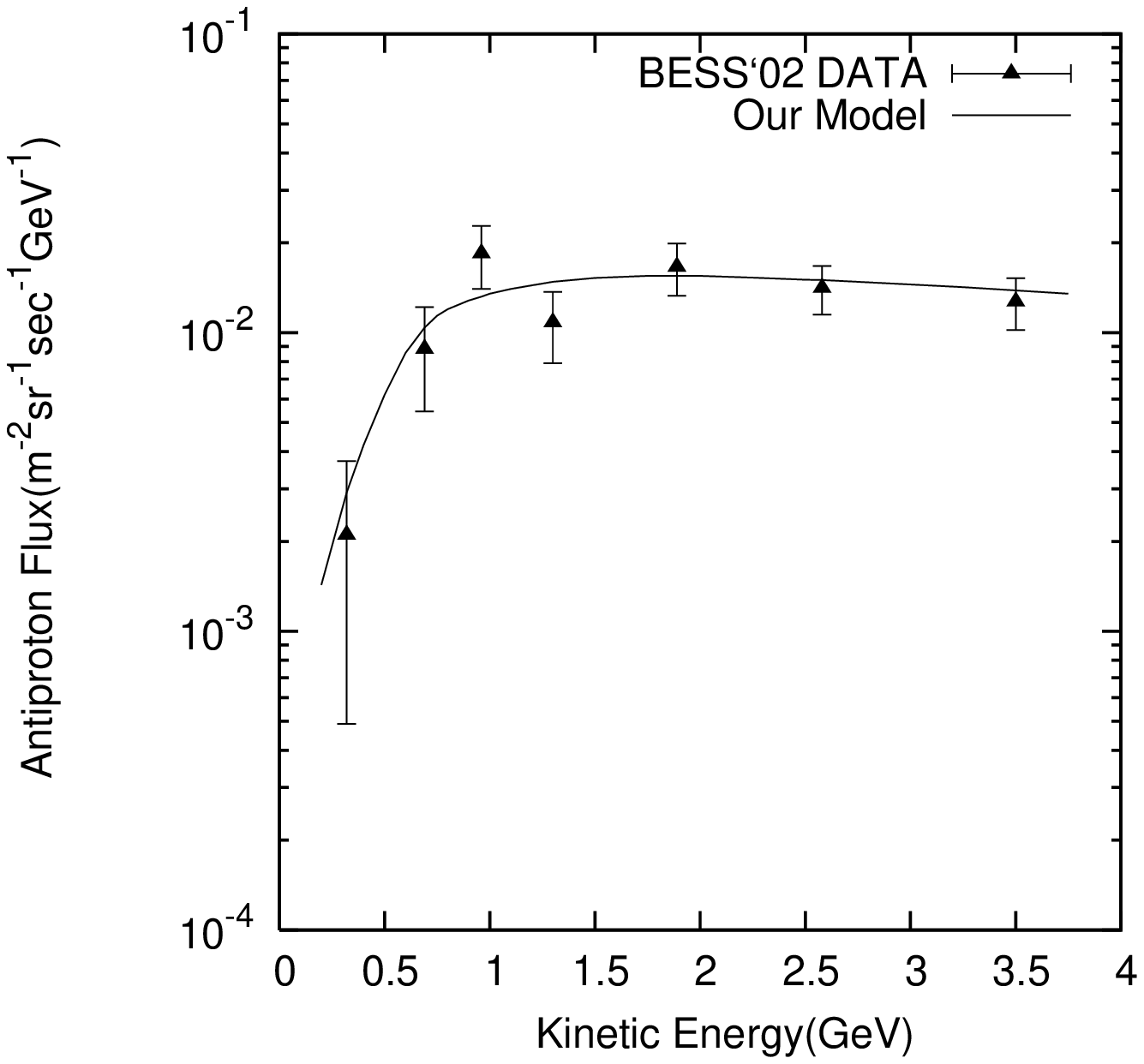}
\end{minipage}}%
\subfigure[]{
\begin{minipage}{.5\textwidth}
\centering
 \includegraphics[width=3.0in]{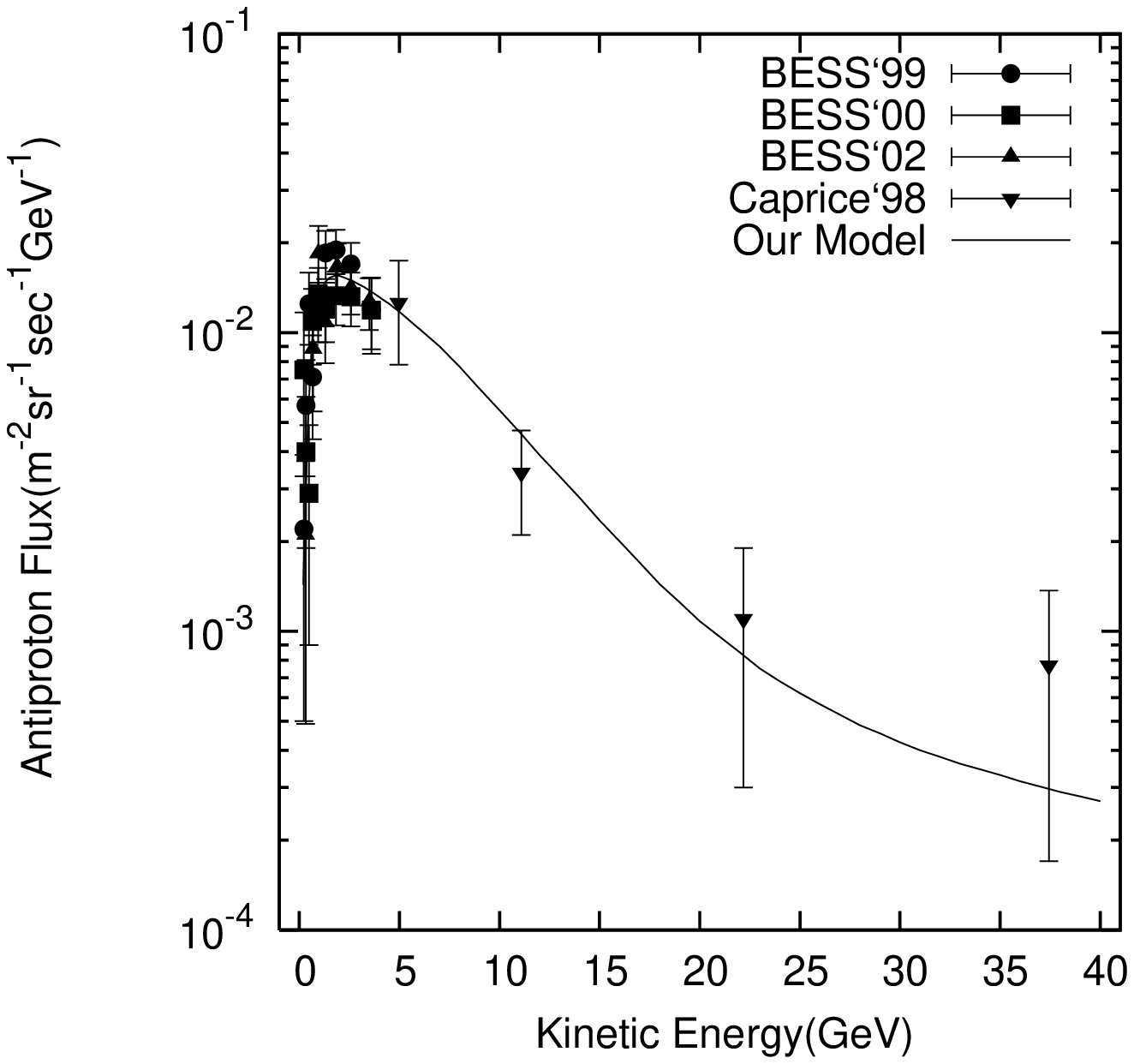}
 \end{minipage}}%
\caption{BESS`99[3(a)], BESS`00[3(b)], BESS`02[3(c)] and together of BESS, CAPRICE[3(d)]
Antiproton fluxes at the top of the atmosphere. The solid curves shows the calculations
of our theoretical model (10). The experimental data are collected from Ref.\cite{Asaoka1,Haino1,Boezio1}}
\end{figure}

\begin{figure}
\subfigure[]{
\begin{minipage}{.5\textwidth}
\centering
\includegraphics[width=3.0in]{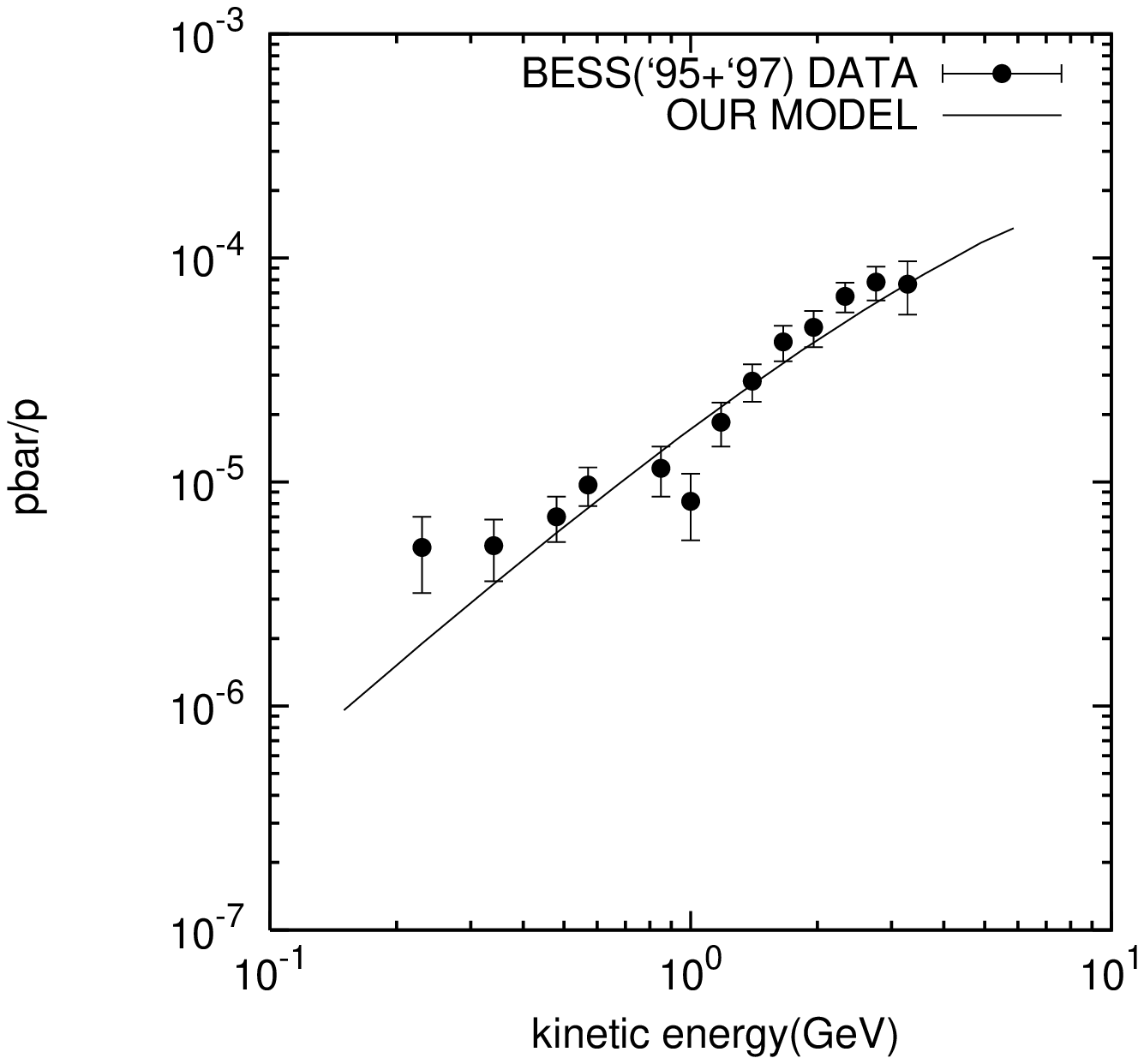}
\setcaptionwidth{2.5in}
\end{minipage}}%
\subfigure[]{
\begin{minipage}{0.5\textwidth}
\centering
 \includegraphics[width=3.0in]{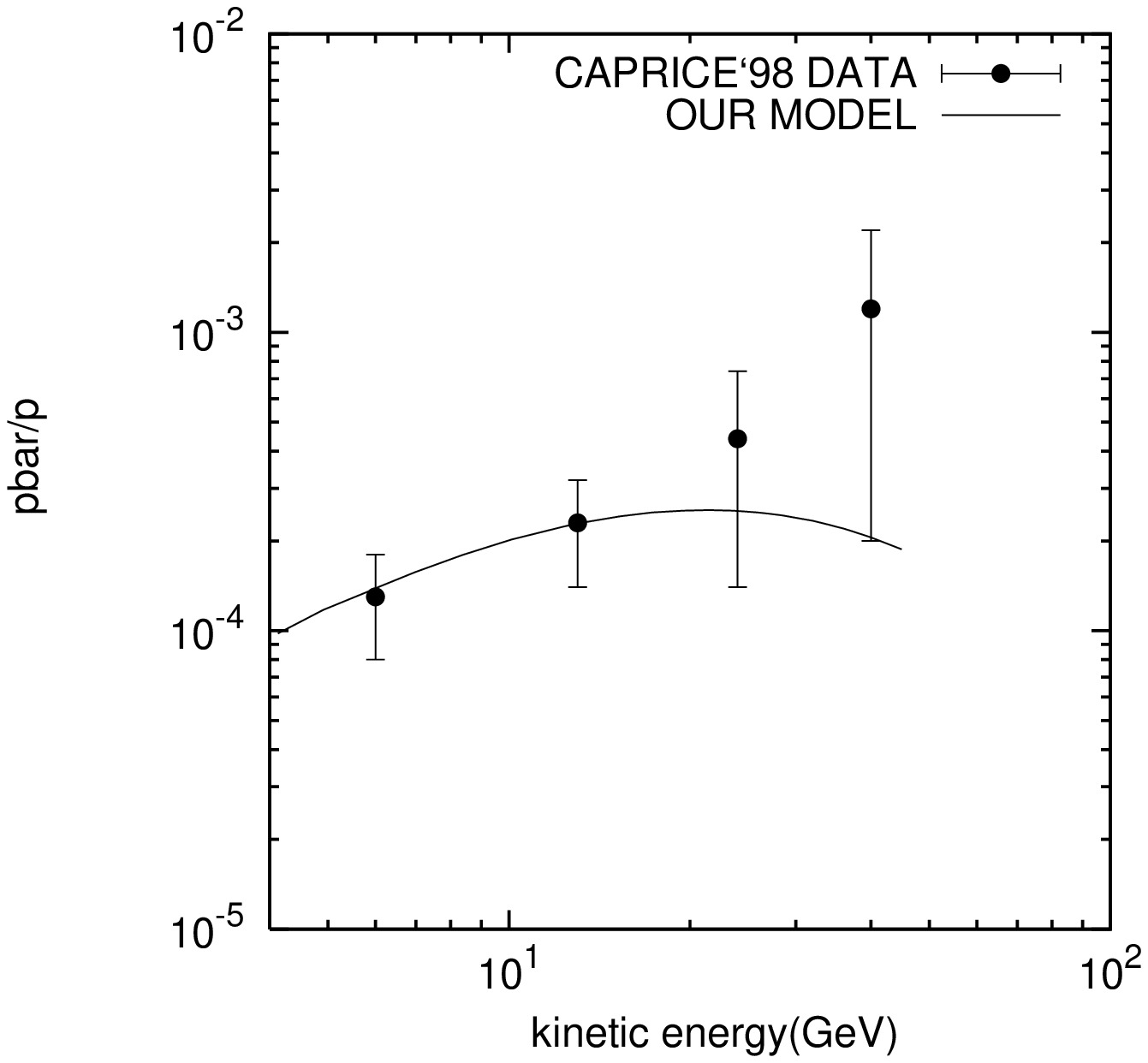}
 \end{minipage}}%
\vspace{0.01in} \subfigure[]{
\begin{minipage}{0.5\textwidth}
\centering
\includegraphics[width=3.0in]{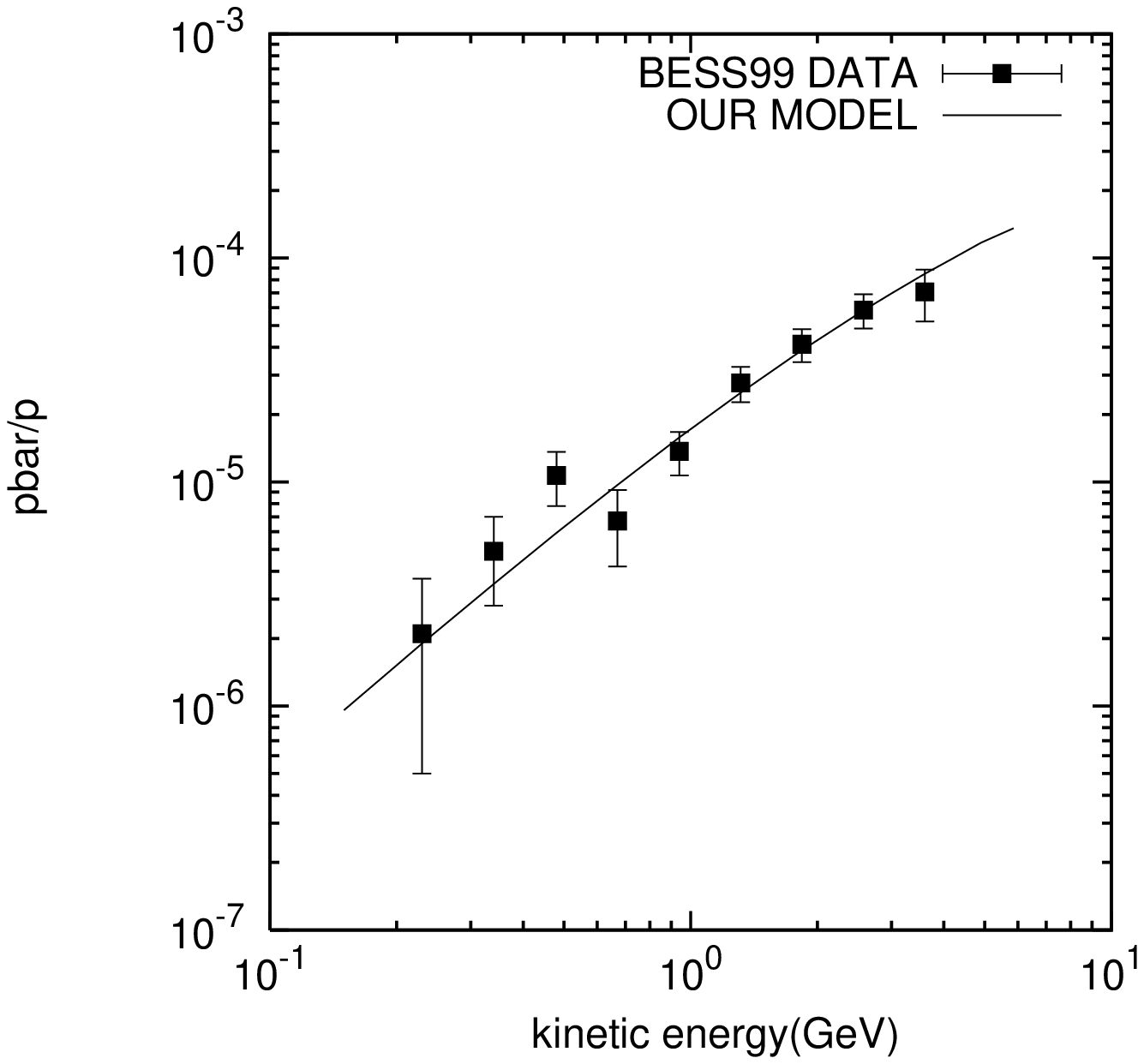}
\end{minipage}}%
\subfigure[]{
\begin{minipage}{.5\textwidth}
\centering
 \includegraphics[width=3.0in]{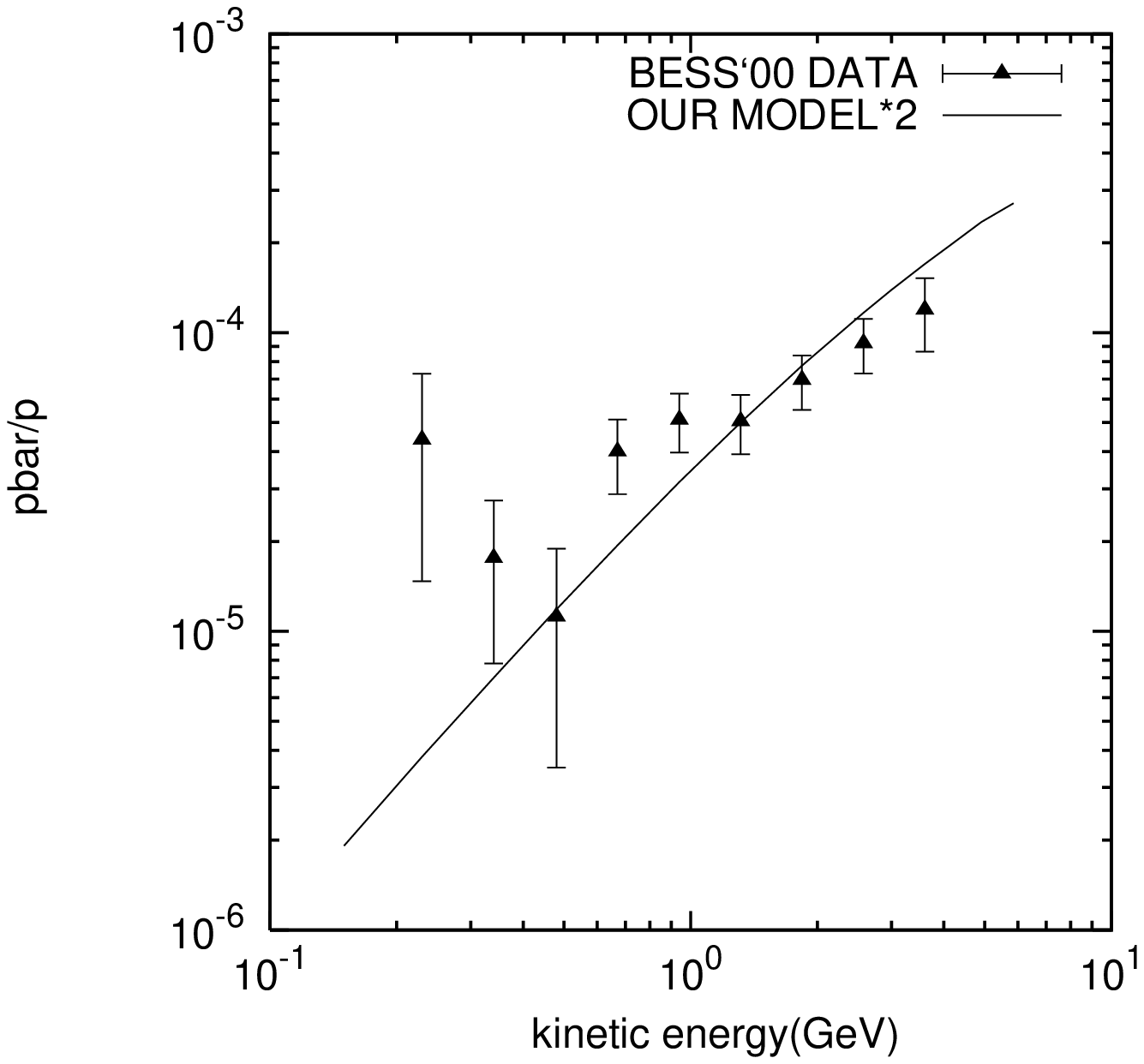}
 \end{minipage}}%
\caption{Plot of $\overline{P}/P$ ratios measured by BESS (`95+`97)[4(a)], CAPRICE`98[4(b)], BESS`99[4(c)] and
BESS`00[4(d)]. The solid curves represent our calculations based on our model based approach (11).
The experimental data are collected from Ref.\cite{Orito1,Boezio1,Asaoka1}}
\end{figure}

\begin{figure}
\subfigure[]{
\begin{minipage}{.5\textwidth}
\centering
\includegraphics[width=3.0in]{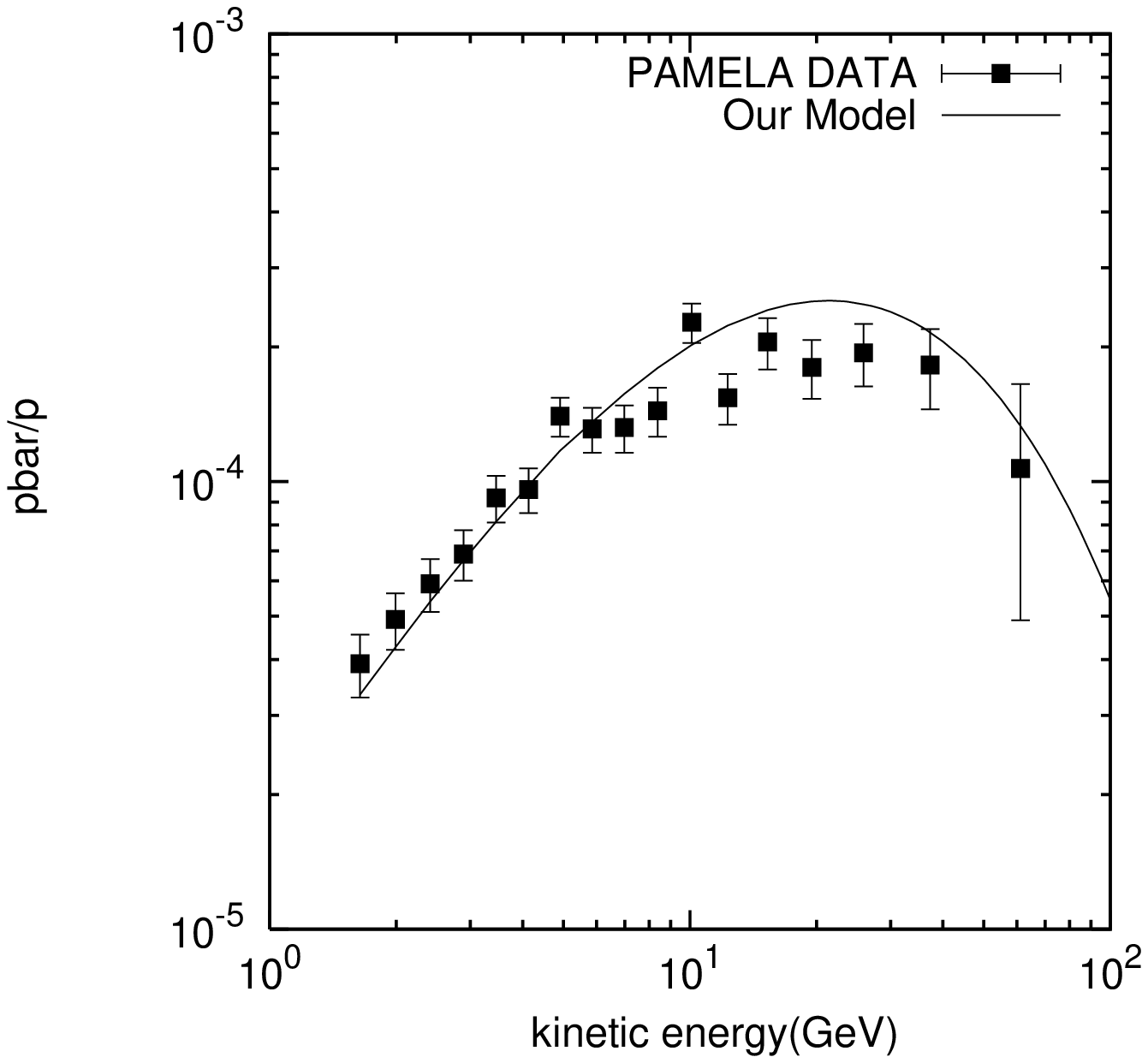}
\end{minipage}}%
\subfigure[]{
\begin{minipage}{.5\textwidth}
\centering
 \includegraphics[width=3.0in]{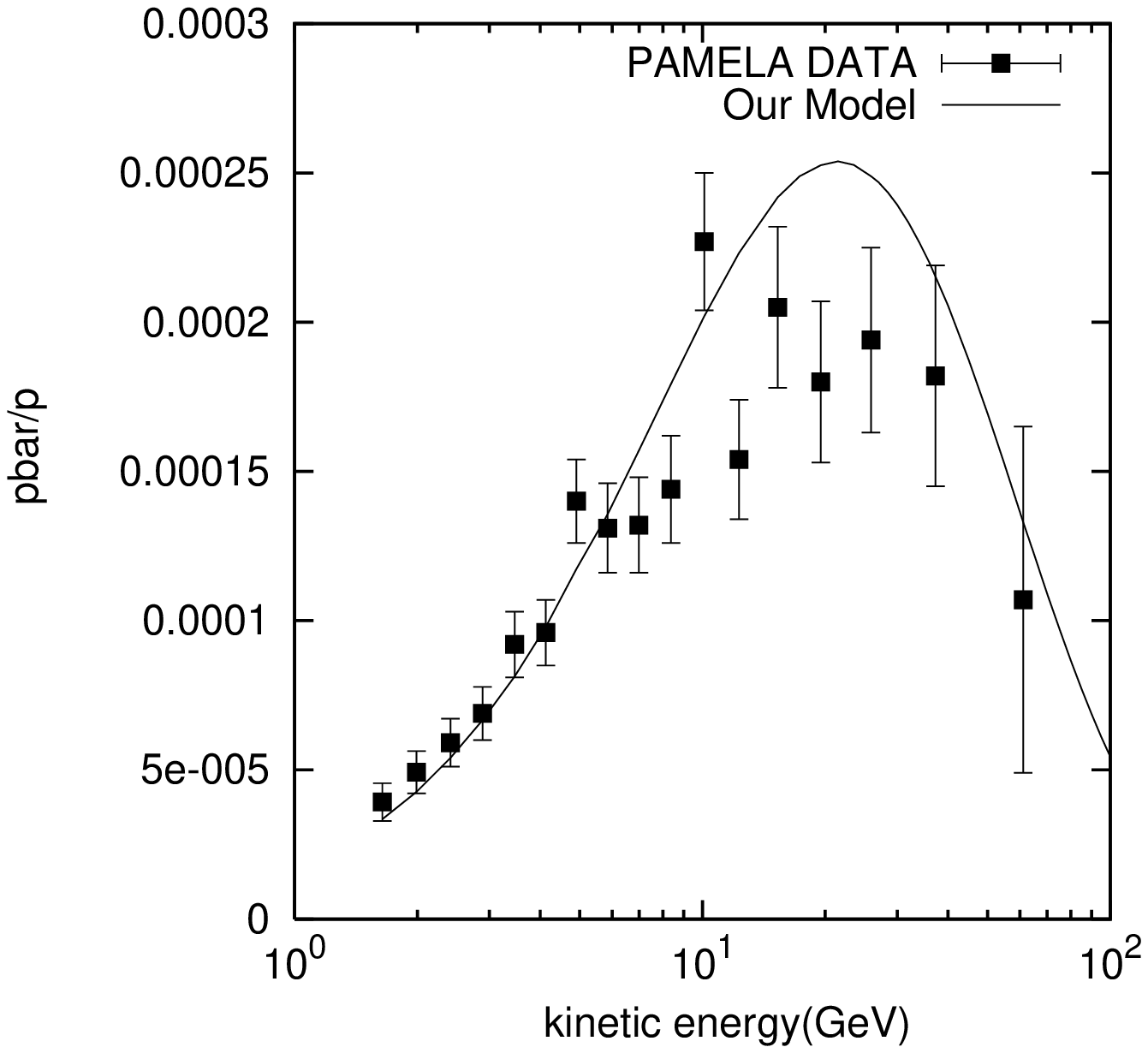}
 \end{minipage}}%
\caption{Plot of $\overline{P}/P$ ratios measured by PAMELA group and the solid curves represent our calculations based
on our model based approach (11). The experimental data are collected from Ref.\cite{Adriani1}}
\end{figure}


\begin{thebibliography}{*}
\bibitem{Adriani1}O. Adriani et al : Phys. Rev. Lett. \textbf{102} (2009)
051101 [astro-ph/0810.4994 v2 25 February 2009].
\bibitem{Adriani2}O. Adriani et al : Nature \textbf{458} (2009) 607-609.
\bibitem{Sau1}Goutam Sau, S.K.Biswas and S.Bhattacharyya : Hadronic
Journal \textbf{31} (2008) 529.
\bibitem{Delahaye1}T. Delahaye, P. Brun, F. Donato, N. Fornengo, J. Lavalle,
R. Lineros, R. Taillet and P. Salati : Proceedings for XLIVemes
rencontres de Moriond, Electroweak Interactions and Unified Theories
session; LAPTH-Conf-1321/09 [hep-ph/0905.2144 v2 20 May 2009].
\bibitem{Dar1} Arnon Dar : Summary talk at the 44th Rencontre De Moriond on High Energy Phenomena
In The Universe which was held in La Thuile, Italy during February
1-8, 2009 [astro-ph.HE/0906.0973 v1 04 June 2009].
\bibitem{Orito1}S. Orito et al : Phys. Rev. Lett. \textbf{84} (2000)
1078-1081 [astro-ph/9906426 v1 26 June 1999].
\bibitem{Asaoka1}Y. Asaoka et al : Phys. Rev. Lett. \textbf{88} (2002)
051101 [astro-ph/0109007 v2 25 January 2002].
\bibitem{Haino1}S. Haino et al : 29th International Cosmic Ray Conference, Pune \textbf{3} (2005) 13-16.
\bibitem{Boezio1}M. Boezio et al : The Astrophysical Journal \textbf{561} (2001) 787.
\bibitem{Cowsik3}R. Cowsik and B. Burch : astro-ph.CO/0908.3494 v1 24
August 2009.
\bibitem{Bandyopadhyay1}P. Bandyopadhyay and S. Bhattacharyya : Nuovo Cimento
\textbf{A, 43} (1978) 323.
\bibitem{Bandyopadhyay2}P. Bandyopadhyay, R. K. Roy Chowdhury, S. Bhattacharyya
and D. P. Bhattacharyya : ~~~~~~~~~~~~~~~ Nuovo Cimento \textbf{A,
50} (1979) 133.
\bibitem{Chou1}T. T. Chou, C. N. Yang and E. Yen : Phys. Rev. Lett.
\textbf{54} (1985) 510.
\bibitem{Alner1}Bonn-Brussels-Cambridge-CERN-Stockholm-UA5 Collaboration
(G. J. Alner et al) : ~~~~~~~~~~~~~~~ Nucl. Phys. \textbf{B, 258}
(1985) 505.
\bibitem{Tan1}L. C. Tan and L. K. Ng : J. Phys. \textbf{G, 7} (1981)
123.
\bibitem{Bhattacharyya1}S. Bhattacharyya and P. Pal : IL Nuovo Cimento \textbf{C,
9} (1986) 961.
\bibitem{Bhattacharyya2} S. Bhattacharyya and D. Roy : Mod. Phys. Letts. \textbf{A,
13} (1998) 2173.
\bibitem{Badhwar1}G. D. Bhadwar, S. A. Stephens and R. L. Golden : Phys.
Rev. \textbf{D, 15} (1977) 820.
\bibitem{Blasi1}P. Blasi : Phys. Rev. Lett. \textbf{103} (2009)
051104 [astro-ph.HE/0903.2794 v1 16 March 2009].
\bibitem{Blasi2} P. Blasi and P. D. Serpico : Phys. Rev. Lett.
\textbf{103} (2009) 081103 [astro-ph.HE/0904.0871 v2 30 September
2009].
\bibitem{Ahlers1}M. Ahlers, P. Mertsch and S. Sarkar :
astro-ph.HE/0909.4060 v1 22 September 2009.
\bibitem{Mertsch1}P. Mertsch and S. Sarkar : Phys. Rev. Lett. \textbf{103}
(2009) 081104 [astro-ph.HE/0905.3152 v3 30 July 2009].
\bibitem{http1}http://ams.cern.ch/
\bibitem{Lineros1}Roberto Lineros : Proceedings for conference "Topics in Astroparticle and
Underground Physics" (TAUP2009) Rome, July 1-5, 2009  [astro-ph.GA/0910.2671 v1 14 October
2009].
\bibitem{Buchmuller1}W. Buchm$\ddot{u}$ller, A. Ibarra, T. Shindou, F. Takayama and D.
Tran :Journal of Cosmology and Astroparticle Physics, 021 (2009)
0909 [hep-ph/0906.1187 v3 31 August 2009].
\bibitem{Cowsik1}R. Cowsik and B. Burch : astro-ph.CO/0905.2136 v2
11 June 2009.
\bibitem{Cowsik2}R. Cowsik and B. Burch :To appear in Proceedings of the 31st ICRC [astro-ph.CO/0906.2365 v1 12 June
2009].


\end{thebibliography}
\end{document}